\documentclass[preprint,preprintnumbers,amsmath,amssymb]{revtex4}

\usepackage{graphicx}

\begin{document}
\draft
\title{Possible evidence for a $5.86 PeV$ cosmic ray enhancement}

\author{ Robert Ehrlich}
\address{George Mason University, Fairfax, VA 22030, USA}
\email{rehrlich@gmu.edu}
\date{\today}

\begin{abstract}

A blind search has been made for cosmic ray sources of neutral hadrons yielding a peak just above the knee, which has resulted in possible evidence for a peak at $5.86\pm 0.75 PeV.$  This search was motivated by a 1999 claim by this author of such a peak at $4.5\pm 2.2 PeV,$ and also some recent results by at least three experiments showing a $E\approx 5.6  PeV$ peak in the all-particle cosmic ray spectrum.   
\end{abstract}
\maketitle
\section{Introduction}

A blind search has been made for cosmic ray sources of neutral hadrons yielding a peak just above the knee.   We report support for such a possibility, i.e., a peak at $5.86\pm 0.75 PeV,$ using data from an international collaboration (Tunka).\citep{Tunka,Tunka1,tunka1}   This search was motivated by a 1999 claim by this author of such a peak at $4.5\pm 2.2 PeV,$\citep{Ehrlich2} and also some recent results by at least three experiments showing a $E\approx 5.6  PeV$ peak in the all-particle cosmic ray spectrum.\citep{rio_conf}  Moreover, its existence had been suggested before the first 1999 claim, based on a model\citep{Ehrlich1} that fit the cosmic ray spectrum, assuming the knee is the threshold for proton beta decay -- a process that becomes energetically allowed if the electron neutrino were a tachyon of mass $m^2\approx -0.25eV^2,$ based on an idea proposed by Chodos, Hauser, and Kostelecky.\citep{Chodos} As noted in ref.~\citep{Ehrlich2} the threshold in PeV for proton decay can be written as: $E_{th}\approx 1.7/\sqrt{-m_{\nu}^2}$

Furthermore, apart from the peak itself, it is noteworthy that the 1999 cosmic ray spectrum fit (Fig. 1 of ref.~\citep{Ehrlich1}) showed significant oscillations in the region above the knee that match those in a recent report of fine structure in this energy region (see Fig. 1 of ref.~\citep{rio_conf}).  Regretably, that 1999 model did err for the region $E > 10^{20} eV$ in failing to include a GZK cut-off.\citep{GZK} Despite that flaw, the inclusion of a GZK cut-off was not a significant feature of the model.  In fact, had one merely assumed a greater distance for the extragalactic sources that would have accommodated a GZK cut-off without affecting the fit for $E < 10^{20} eV.$ The one essential feature of the 1999 model was that cosmic ray protons began to decay when $E>E_{knee},$ resulting in a decay chain: $p\rightarrow n \rightarrow p\rightarrow n \rightarrow \cdots$ Such a decay chain would continue until the baryon's energy drops below $E_{knee},$ thereby giving rise to the knee, and resulting in a pile-up of neutrons just above it, i.e., a peak at  $E = 4.5 \pm 2.2 PeV.$\citep{Ehrlich1} Neutrons, being uncharged, mostly point back to their sources, unlike protons whose directions are randomized by the galactic magnetic field. Thus, if the baryon in the decay chain spends most of its time as a neutron most of its directional information should be preserved en route to Earth.  Moreover, the hypothesized decay chain could allow PeV neutrons to reach us from sources normally considered too distant, given the neutron lifetime such as Cygnus X-3.

The first claim\citep{Ehrlich2} for a $4.5 PeV$ peak was based on Lloyd-Evans data for Cygnus X-3.\citep{Lloyd-Evans}  As a binary having an orbital period of 4.79 h, for certain rotation phases jets emanating along the rotation axis of one member of the binary might point towards Earth and increase the observed signal, which is what the Lloyd-Evans data seemed to show. In fact, the reported signal was seen by selecting events in a particular $2.5\%$ wide phase window, and using as background the remaining $97.5\%$.  If the signal was real, such a phase window cut would reduce background by a factor as much as 40, greatly enhancing the signal.  Using the Lloyd-Evans data Ehrlich showed that the two bins straddling $5 PeV$ had $28.4\pm 4.7$ excess events ($6.0\sigma$).\citep{Ehrlich2} Apart from skepticism of this claim, there is also much skepticism about the existence of Cygnus X-3 as a source of PeV cosmic rays. However, the basis of that skepticism may be poorly justified, especially if Cygnus X-3 is an episodic source, and if a weak $E=4.5 PeV$ signal needs cuts to suppress background, as discussed in more detail in Appendix I. 

\section{The Tunka experiment}

Recent support for a cosmic ray peak just above the spectrum knee can be found in data reported by the Tunka collaboration, even though the authors characterize their observation instead as an example of a ``remarkable fine structure" seen above the knee in the all-particle cosmic ray spectrum.\citep{rio_conf}  Nevertheless, Fig. 1 of their paper does show an  unambiguous peak at $E\approx 5.6 PeV$ in the combined Tunka-25 and Tunka-133 data.  One might well be suspicious in any peak that occurs just at an energy where the spectra from the two data sets join  On the other hand, it is quite significant that Fig. 1 of their paper shows data from three other experiments that exhibit the same peak as Tunka (KASCADE Grande, Ice Top, and Tibet).\citep{Kascade,Ice_top,Tibet1}  The Tunka authers interpret the fine structure above the knee as being quite consistent with a combined source model (galactic SN remnants plus extragalactic source(s)), with a suitable choice of free parameters.  However, the authors do not consider an alternative hypotheses that can also account for the observed ``remarkable fine structure" (including the $E\approx 5.6 PeV$ peak) -- in particular they do not mention the 1999 prediction and subsequent observation of just such a peak previously predicted and then observed at $4.5 \pm 2.2 PeV.$  In what follows, we present an independent analysis of the Tunka-133 data, which seeks corroborating evidence for such a peak by attempting to find possible cosmic ray sources in a blind all-sky survey.  We have no knowledge of the position of Dr. Kuzmichev, head of Tunka, or other members of the collaboration on the results presented here.  

\subsection{Tunka: history, efficiency, and exposure}

Tunka began in the 1990's and it observes the extensive air showers produced by cosmic rays.  Originally Tunka operated using 25 Cherenkov counters, but the newer Tunka-133 array used 133 Cherenkov counters covering an area of $1 km^2.$  The Tunka-133 data analyzed here was collected during three successive winter seasons from 2009 to 2012 during clear moonless nights. It consists of approximately $1.8\times10^6$ events with zenith angle less than $45^0$ and energies $E>1PeV$ measured to an precision of $15\%.$\citep{Tunka}  The efficiency of the detectors as a function of zenith is close to $100\%$ up to $30^0$ and reduces to $50\%$ at zenith angles above $45^0$. As a function of energy, the efficiency for $E > 6 PeV$ is $100\%$ and the threshold is $1 PeV,$ so at $E=4.5 PeV$ it is above $50\%.$ Tunka-133 also has good exposure in the Northern Hemisphere, with a field of view of $\pi$ steradians.\citep{Exposure}  

\subsection{Data analysis}

The only cut made on the Tunka-133 data set was that the zenith angle should be less than $45^0,$ above which the acceptance drops below $50\%.$   An energy histogram of the whole data set using energy intervals of width $\Delta Log_{10} E =0.1$ shows about 280,000 events in the peak energy bin straddling $2.8 PeV.$  In order to search for evidence for a peak near $E \approx 5.6 PeV$ we do not focus on any specific possible source such as Cygnus X-3, but instead first identify a large number of ``candidate sources," defined arbitrarily as small non-overlapping circular regions of the sky having $S > 3.3\sigma$ excess counts above background in each energy bin, and then examine the energy distribution of these candidate sources in five energy bins.   

The statistical significance S is found in terms of the on-source counts $n_{on}$ and the background counts $n_{Bkd}$ using:
\begin{equation}
S=\frac{n_{on}- n_{Bkd}}{\sqrt{n_{Bkd}}}
\end{equation}
In Eq. 1, the background count is found using the shuffling method,\citep{Cassiday} which involves generating artificial events by shuffling the arrival times of all events having very close altitude and azimuth coordinates to generate a set of artificial events.  In the absence of any real sources, this method allows one to calculate an accurate background in celestial coordinates (right ascension and declination).  The systematic error using the shuffling method is less than 0.0008 events for $0.2^0 \times 0.2^0$ bins, based on 10,000 cycles of shuffling. Thus, for the largest radius search window used ($4.5^0$) one would expect a systematic error of less than 1.3 counts, which is negligible compared to the statistical fluctuations. 

The search procedure is to scan the sky in right ascension $\alpha$ and declination $\delta,$ and calculate $S$ based on the number of events above background in each small non-overlapping circular region.  The procedure is then repeated many times, each time shifting the search pattern of circles by $1^0$ in $\alpha$ or $\delta$.  Note that in the model in which the $4.5 \pm 2.2 PeV$ peak was predicted, there was no mention of the angular spread of neutrons arriving from cosmic ray sources, because that would depend on proton decay dynamics and assumptions on the distribution of source distances. Therefore, when doing a blind search it is reasonable to use multiple search radii, and the following three are used: $r=1.5^0,$ $3.0^0,$ and $4.5^0.$    No other search radii were tried.  This choice of the three radii was based on the following considerations: (a) the smallest window used is large enough compared to the angular resolution\citep{tunka1} to give meaningful results, (b) the largest one equals the largest window for which signals had earlier been claimed for Cygnus X-3.  Finally, (c) one intermediate choice between them was considered sufficiently different from the others that one might expect to find many possible candidate sources not picked up by the other two.

Let us define $n(r, S,E)$ as the number of times we find an excess above background at a significance level S using a search radius $r$ for cosmic rays having for the energy bin centered on energy $E$ (in PeV), and we also define: 
\begin{equation}
N(S,E)=n(1.5^0, S,E) +n(3.0^0, S,E) +n(4.5^0, S,E)
\end{equation}
as the number of times we find an excess above background at a significance level S using any of the three search radii.  We only use cases, however, where each $n(r, S, E)>50,$ so as to avoid spurious large values of S resulting from very small numbers, and to avoid a breakdown in the Gaussian $\sqrt{n}$ approximation of errors.

\subsection{Results}

We have examined evidence for a peak near the one at $E\approx 5.6 PeV$ seen by Tunka and three other experiments in the all-particle cosmic ray spectrum.  Note, however, that an all-sky spectrum of Tunka-133 data (not combined with Tunka-25) shows no hint of a peak at $E\approx 5.6 PeV;$ it is only when searching for numbers of candidate sources that one appears.  Specifically, when we use five bins of width $\Delta Log_{10} E = 0.1$ centered on $5.86 PeV,$ it is that central energy bin which shows the greatest excess number of candidate sources above what chance predicts.  A peak at this energy is quite consistent with both the all-particle spectrum peak at $E \approx 5.6 PeV,$ as well as the earlier reported peak at $E = 4.5\pm 2.2 PeV.$  Fig. 1(a) shows a histogram of $N(S,5.86)$ versus S, i.e., the number of times various positive and negative $S$ values are found in the all-sky scan for the energy bin centered on $E=5.86 PeV.$   As can be seen, the data agree quite well with the expected $\sigma=1$ Gaussian distribution for $S<0,$ but for $S >0$ the departure from the Gaussian appears more pronounced at large $S.$  For example, while there are 68 candidate sources, i.e., $S > +3.3$ regions, there are only 18 cases where $S < -3.3 ,$ which is very close to what the Gaussian distribution predicts, i.e., 20.  For greater clarity in showing the magnitude of the deviation from the Gaussian for large $S >0,$ Fig 1 (b) shows a blow-up of Fig 1(a) for the region $S > 3.0.$

Most interestingly, this very pronounced deviation from a Gaussian for large positive $S$ is seen ${\emph only}$ when the S-distribution is examined for the energy bin centered on $E=5.86 PeV,$ and it is significantly less at smaller and larger energies -- see Fig 2(a), which shows the number of candidate sources as a function of energy in excess of the 20 that the Gaussian distribution predicts for the only five energy bins we have examined.  It must be emphasized that it is the $\emph{excess}$ number of candidate sources above background that is plotted in Fig 2(a), so that one expects the five data points in the absence of real sources to be consistent with zero at all energies, not with an arbitrary horizontal line.  Thus, in contrast, to the systematic effect seen for the candidate sources, Fig. 2(b) shows that the number of excess candidate ``sinks", i.e., $S < -3.3 $ regions in fact shows no statistically significant departures from zero for all energy bins.  

\subsection{The oversampling bias}

The over-sampling the same regions of the sky in our search procedure can inflate greatly the statistical significance of the number of candidate sources. Thus, based on Fig. 2(a) it would {\emph not} be valid to conclude that the $\emph excess$ number of candidate sources for the $E=5.86 PeV$ bin is $68 -20 = 48 \pm \sqrt 20.$   The oversampling bias can be removed only partly by enlarging the uncertainty on the number of sources using 3 search windows by a factor of $\sqrt{3},$ yielding for the number of excess candidate sources above chance to $48 \pm 7.8.$  This penalty factor does not go nearly far enough, however, because the biggest source of oversampling (especially for the largest size search window) results from the fact that a localized excess or deficiency having a size comparable to the search window might be picked up by many nearby windows of that size, as we step the whole grid in $1^0$ steps in $\alpha$ and $\delta$ across the sky.  A partial way to remove this source of oversampling is to use the distribution in the number of candidate sinks to estimate the uncertainty in the number that chance would predict.  In other words, since no real sinks are physically possible their rms deviations from zero serves as a good estimate of the uncertainty in the excess numbers of sources and sinks.  We find using the five energy bins, an rms deviation of 12, which how the size of the error bars in Figs. 2(a) and 2(b) were determined.  Combining the excess counts for the bin centered on $5.86 PeV$ with the two adjacent bins, one finds an excess of 102.7 with an uncertainty of $12\sqrt{3}.$   Were we confident that the oversampling has been removed based on this analysis, the result could be expressed as $102.7\pm 20.8 (5.0 \sigma)$ excess candidate sources above chance, $\emph{but no such claim can be made here}$, and hence our result remains to be corroborated by other data sets.  

Fig. 3 shows the locations (right ascension $\alpha$ and declination $\delta$) of the 68 candidate sources (upper graph) and $1/4$ as numerous sinks (lower graph) for the $5.86 PeV$ energy bin.   It is clear that the largest fraction of the candidate sources in this energy bin have the $4.5^0$ radius.  The concentrations of candidate sources in specific $\alpha, \delta$ regions could in principle have a physical basis, but it also is undoubtedly due to the previously discussed over-sampling bias, especially for the concentration in the vicinity of $\alpha=10^0$ and $\delta=15^0$ where the degree of concentration is most extreme.  In the other three concentrations near $(\alpha, \delta) = (260^0, 15^0), (260^0, 50^0), (10^0, 50^0),$  the spacings between candidate sources is sufficiently large that real physical concentrations of sources is a realistic possibility, even if some amount of oversampling also exists. 

\subsection{Summary}

In summary, support for a peak at $5.86 \pm 0.75 PeV$ is presented -- a value consistent with (a) the $E\approx 5.6 PeV$ peak seen in the all-particle spectra from Tunka and three other groups, (b) the 1999 claim of a peak at $E=4.5\pm 2.2 Pev$ reported for Cygnus X-3 data, and (c) the prior 1999 prediction of such a peak at this energy.  The new evidence is $\emph {not}$ in the form of excess counts above what background for any suspected identifiable source(s), but rather an excessive number of candidate sources, i.e., small circular regions of the sky having at least $3.3\sigma$ excess counts above chance in  energy bins near $5.86 PeV$ -- a procedure that has the effect of greatly enhancing any weak signal, assuming that there are numerous real sources at unknown locations.  Moreover the numbers of candidate sources seen versus energy falls as one moves away from $5.86 PeV$ in either direction, as expected, and no comparable excess is seen in the number of candidate ``sinks" for any of the five energy bins examined.  Although Fig. 2(b) might suggest a hint of an energy dependence for the numbers of excess sinks above chance, the chi square probability for a fit to a horizontal straight line at zero height $(29\%)$ indicates consistency with the previous assertion.  Despite the possible evidence for a peak at $5.86 \pm 0.75 PeV$ presented, it is recognized that in view of the methodology used, i.e., searching for sources at locations where no objects are known to exist, and a failure to deal completely with the oversampling bias, our result can only serve as a motivation to others to see if there is evidence for the enhancement claimed for sources in the same regions of the sky that have been identified here.

If the peak is real, one reason it may not have been claimed previously by others is that most experimenters who saw no statistically significant excess of cosmic rays from the direction of any suspected source probably would have little reason to look at any particular energy band.  Obviously, the peak identified here will need to be seen in the new data collected by Tunka and/or by others having data sets with sufficient numbers of events near the knee before it will be regarded seriously. Any positive result from future searches would be quite interesting -- $\emph{particularly}$ if the locations of most of the candidate sources match those in Fig. 3 (a).

\begin{acknowledgments}
I am immensely grateful to Dr. Kuzmichev for providing access to the Tunka data, and to Dr. Mikhail Zotov for providing his analysis of the data, and to both of them for supplying specific requested information about the Tunka data.  
\end{acknowledgments}

\section{Appendix I: Skepticism about Cygnus X-3 data}
The Cygnus X-3 observations from the 1980's are now considered highly suspect by many cosmic ray researchers, in light of a subsequent negative searches in which the CASA-MIA and EAS-TOP Collaborations reported no non-episodic signals from Cygnus X-3 or anywhere else.\citep{Borione, Aglietta} Nevertheless, the earlier experiments could be correct, since Cygnus X-3 is known to be a highly episodic source, and in fact during strong flares its radio emissions increase a thousand fold, so it is possible that the source might simply have gone quiet during most of the periods those experiments were in operation. A data set making this possibility plausible was taken by the Tibet AS-gamma experiment, in which a strong signal was seen for the period 1997-2001 in a $3^0$ circle around Cygnus X-3, but virtually no signal for the subsequent 4-year period.\citep{Tibet} In addition, the negative result in the CASA-MIA experiment says little about the reality of a 4.5 PeV peak, since the number of events for $E>1.17 PeV$ for CASA-MIA was a mere 0.1\% of the total and that for $E>E_{knee}$ was negligible. 

Moreover, other data sets exist which are consistent with the 1980's Cygnus X-3 claims. In addition to the Tibet AS-gamma experiment, Marshak et al. observed anomalous deep underground muons over a ten year period within $2^0$ of Cygnus X-3.\citep{Marshak} The existence of deep underground muons seen in that experiment implied that the primary particles producing them were strongly interacting, and the pointing back to Cygnus X-3, required them to be neutral, i.e. neutrons or some new exotic hadrons. Even the size of a $2^0$ circle (significantly more than the experimental resolution) has a ready explanation given the slight loss of directionality of the neutrons implied by the $p\rightarrow n \rightarrow p\rightarrow n \rightarrow p\rightarrow n \rightarrow \cdots$ decay chain. As an indication of the importance of selecting periods of high luminosity, the Marshak et al. signal was seen only for those times during which strong radio flares occurred.\citep{Marshak} The NUSEX experiment also reported excess muons for a $4.5^0$ cone centered on Cygnus X-3.\citep{NUSEX} Unfortunately, in recent years there have been few periods of major flares from Cygnus X-3.  Although there was a brief one in 2011 associated with gamma rays,\citep{Corbel} the last one before that was in 2006.   Likewise, other groups have also observed the Cygnus region as a transient source of high energy gamma rays in recent years,\citep{AGILE, Ackerman} but this of course could not account for underground muons seen in earlier experiments. A true test of the claim of a weak $4.5 PeV$ signal due to neutrons coming from Cygnus X-3 would  require sufficient numbers of events in that energy region, an exposure time that included some major flares, and a selection on the appropriate phase interval, using a precise ephemeris, which is extremely important for a long observing time.\citep{vanderklis}

\begin{figure}[h]
\includegraphics[width=100mm]{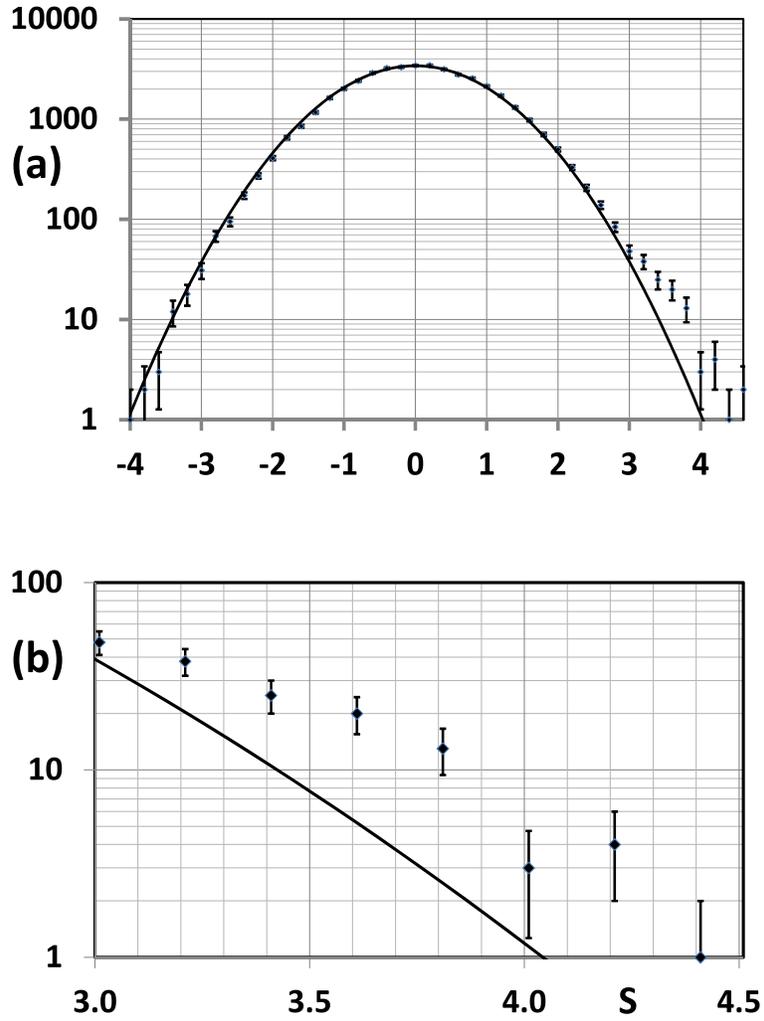}
\caption{$N(S)$ versus S, i.e., numbers of $1.5^0$, $3^0$, and $4.5^0$ regions centered on $1^0$ grid points in an all sky survey showing various levels of statistical significance above background (S in $\sigma$) for the energy bin centered on $E=5.86 PeV.$ The curve is a Gaussian having $\sigma=1.$  For greater clarity fig. 1(b) shows a blow-up of the region above $S>3.0.$} 
\end{figure}

\begin{figure}[h]
\includegraphics[width=100mm]{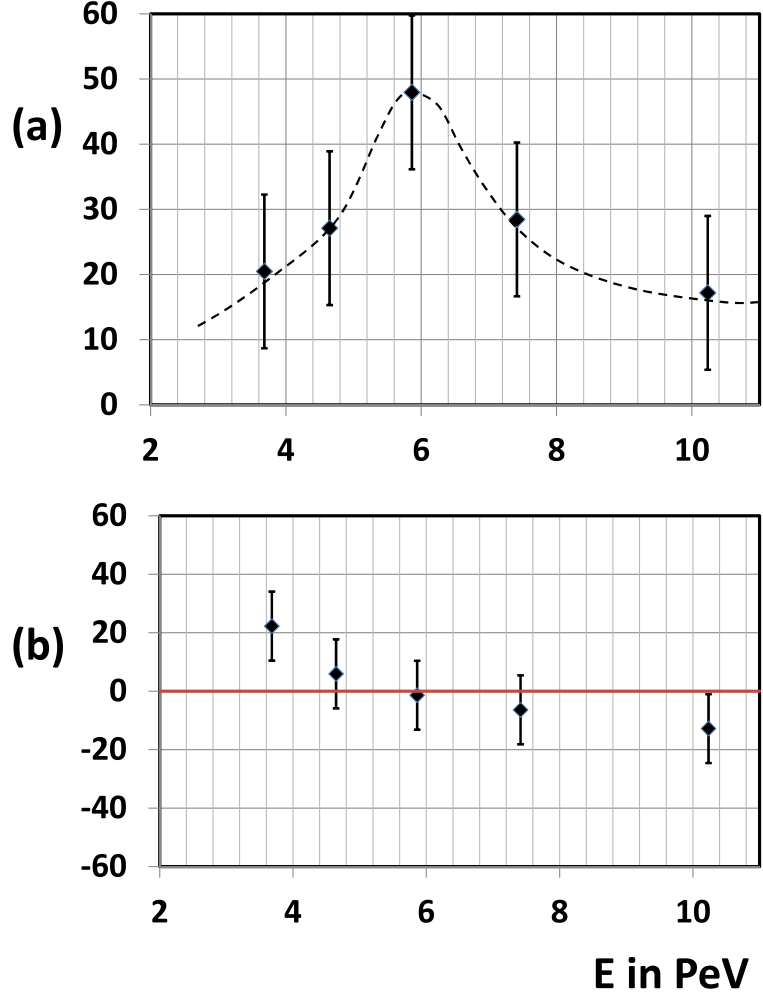}
\caption{The number of ``excess" candidate sources, i.e., above the number the Gaussian predicts, versus energy using bins of width $\Delta Log E = 0.1$.  (a) is for $S > +3.3$, and (b) is for $S < -3.3 $, i.e, ``sinks" rather than sources.  In finding the numbers of candidate sources or sinks a $\pm 2\%$ adjustment was made for each energy bin, based on the varying heights of the best-fit Gaussians, whose normalizations vary from one energy bin to another at the level of $\pm 2\%$.  The dashed curve in (a) is simply a smooth curve through the data points.}
\end{figure}

\begin{figure}[h]
\includegraphics[width=100mm]{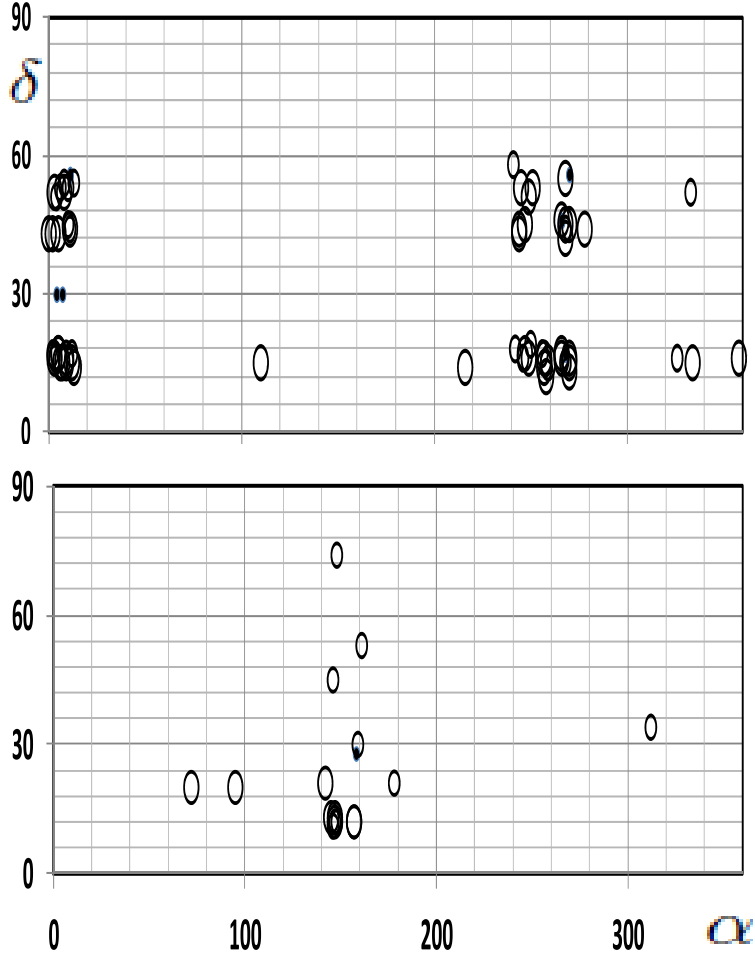}
\caption{Locations in the sky (right ascension $\alpha$, and declination $\delta$) for candidate sources and sinks in the $5.86 PeV$ energy bin: (a) Upper graph is for sources: $S > +3.3,$ and (b) Lower graph is for the far less numerous sinks: $S < -3.3$. The small black ellipses correspond to a search radius of $1.5^0,$ while the medium and large open ellipses indicate search radii of $3.0^0$ or $4.5^0$ respectively.  The use of ellipses is based on the different axis scales.} 
\end{figure}

\end{document}